\documentclass[%
 reprint,
 amsmath,amssymb,
 aps,
]{revtex4-1}

\usepackage{graphicx}
\usepackage{dcolumn}
\usepackage{bm}
\usepackage{upgreek}

\usepackage{amsmath}
\usepackage{textgreek}
\usepackage{amssymb}

\begin{document}

\preprint{APS/123-QED}

\title{Resource-efficient frequency conversion for quantum networks via sequential four-wave mixing}

\author{Thomas A. Wright\textsuperscript{1}}
\email{t.wright@bath.ac.uk}
\author{Charlotte Parry\textsuperscript{1}}%
\author{Oliver R. Gibson\textsuperscript{1}}%
\author{Robert J.A. Francis-Jones\textsuperscript{2}}%
\author{Peter J. Mosley\textsuperscript{1}}%

\affiliation{%
\textsuperscript{1}Centre for Photonics and Photonic Materials, Department of Physics, University of Bath, Bath, BA2 7AY, UK
}%

\affiliation{%
\textsuperscript{2}Clarendon Laboratory, University of Oxford, Parks Road, Oxford, OX1 3PU, UK
}%

\begin{abstract}
We report a resource-efficient scheme in which a single pump laser was used to achieve frequency conversion by Bragg-scattering four-wave mixing in a photonic crystal fiber. We demonstrate bidirectional conversion of coherent light between Sr$^+$ $^2$P$_{1/2}$ $\rightarrow{}$ $^2$D$_{3/2}$ emission wavelength at 1092\,nm and the telecommunication C band with conversion efficiencies of 4.2\,\% and 37\,\% for up- and down-conversion, respectively. We discuss how the scheme may be viably scaled to meet the temporal, spectral and polarisation stability requirements of a hybrid light-matter quantum network. 
\end{abstract}

\maketitle


Quantum networks provide a robust and scalable framework through which large-scale quantum-information processing may be achieved\,\cite{Kimble:2008if}. 
Specifically, the development of hybrid light-matter networks allows a consolidated approach whereby the specific advantages of multifarious quantum platforms are harnessed\,\cite{Walmsley:2016hn,Nigmatullin:2016dx}. For example, trapped ions may be used to achieve high-fidelity gate operations\,\cite{Ballance:2016hy} and fluorescence photons from these trapped ions can transmit information between nodes\,\cite{Bock:2018jw,Stephenson:2020iw}. As ion traps are limited to -- at best -- tens of ions, underpinning the network will be an optical bus in which photonic interconnects distribute entanglement between nodes. Scaling hybrid networks up to larger numbers of nodes or longer distances can only be achieved with optical fiber links.

Unfortunately, many candidate ions emit photons at wavelengths where attenuation in optical fiber is prohibitively high. In order to access the infrared telecommunication bands at which fiber links have minimum loss, photons may be coherently shifted by quantum frequency conversion (QFC) \cite{Huang:1992gf}, usually by three- and four-wave mixing (TWM and FWM) processes. It is therefore key to the  realisation of large-scale hybrid quantum networks to develop QFC techniques to remap photons to and from telecommunication wavelengths.

 The platforms used to achieve QFC are diverse: early experiments used nonlinear crystals \cite{Huang:1992gf,Albota:2004gm,Vandevender:2004ef}, but demonstrations have since broadened to the use of planar waveguides \cite{Langrock:2005ie,Tanzilli:2005jd}, microresonators \cite{Guo:2016iu,Singh:2019di}, optical fibers \cite{McGuinness:2010ja,Clark:2013cia} and atomic systems \cite{Radnaev:2010il,Bustard:2017dh}. 
 Experiments which are directly relevant to quantum networks include the spectral remapping of photons both to \cite{Takesue:2010ba,Ikuta:2011de,Zaske:2012hka,DeGreve:ge, Albrecht:cf,Farrera:2016jt,Krutyanskiy:2017eu,Bock:2018jw,Walker:2018jh,Dreau:2018dn} and from \cite{Tanzilli:2005jd,Vandevender:2007gx,Rakher:2010jua,Rakher:2011jia,Maring:2014dp,Vollmer:2014ia,Baune:2016fd,PhysRevApplied.7.024021,Allgaier:2017hv} infrared (IR) telecommunication bands, but also between dissimilar node wavelengths \cite{Maring:2017eb}. Although two-way conversion is desirable, most reported conversions have been unidirectional with limited exceptions \cite{Dudin:2010ji,Wright:2018ks}.

The majority of reported QFC demonstrations have exploited parametric TWM where the energy of photons in a high-intensity pump field ($\hbar\omega_{\mathrm{P}}$) must equal the energy difference between the input and output photons: $\hbar\omega _{\mathrm{P}} = |\hbar\omega _{\mathrm{in}} - \hbar\omega _{\mathrm{out}}| $. Hence TWM processes lend themselves to large frequency shifts that can be driven by high-power laser systems at convenient wavelengths. In contrast, QFC by Bragg-Scattering FWM (BS-FWM) allows an input photon to be shifted by the frequency difference between two pump fields\,\cite{McKinstrie:2005hja}, enabling smaller frequency shifts where high-power pump lasers for TWM processes may not exist or materials may be opaque. In addition, BS-FWM is rather more flexible as tailoring waveguide dispersion allows the mean wavelength of the pumps to be adjusted and it does not require second-order nonlinearity. Nevertheless the clear disadvantage of BS-FWM is the need to supply two pump fields; a pair of high-power pulsed lasers is often necessary to minimize the nonlinear coupling length, creating a large resource overhead associated with BS-FWM QFC interfaces\,\cite{McGuinness:2010ja}.

\begin{figure}
\centering
{\includegraphics[width=\linewidth]{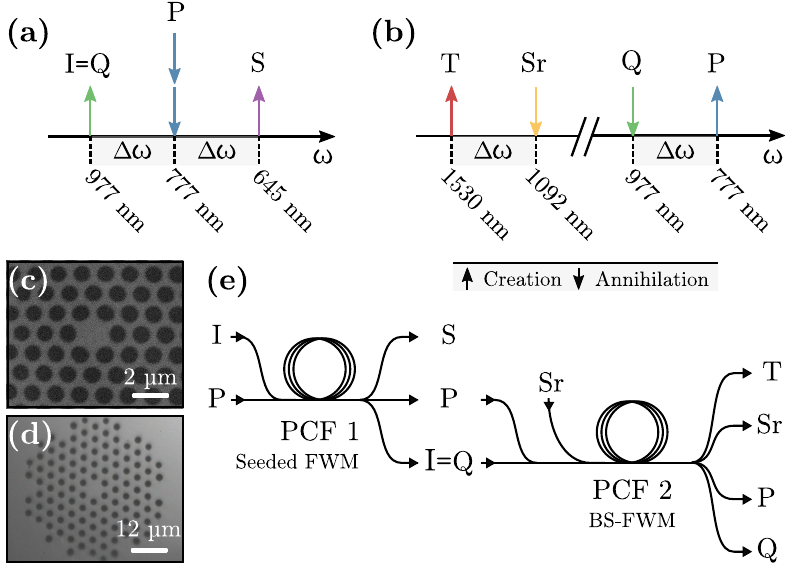}}
\caption{(a) The seeded FWM. (b) The BS-FWM down conversion. (c) Scanning-electron micrograph of PCF\,1 with pitch 1.51\,$\mu$m and hole diameter 0.96\,$\mu$m. (d) Optical micrograph of PCF\,2 with pitch 3.48\,$\mu$m and hole diameter 1.57\,$\mu$m. (e) The conversion scheme: the pump, P, and idler, I, fields from a seeded-FWM process in PCF\,1 are used in PCF\,2 to pump BS-FWM conversion of photons between a Sr$^+$ emission wavelength, Sr, and the C-band wavelength, T. Q denotes use of the idler as a long-wavelength pump. S denotes the signal field.}
\label{fig:1}
\end{figure}

In this letter we present resource-efficient BS-FWM frequency conversion between the Sr$^+$ emission wavelength at 1092\,nm and the telecommunication C band. Sr$^+$ ions are strong candidates for use in quantum networks \cite{Stephenson:2020iw}, with the $^2$P$_{1/2}$ $\rightarrow{}$ $^2$D$_{3/2}$ transition presenting an opportunity to alleviate some engineering complexities associated with the 422\,nm emission from the $^2$S$_{1/2}$ $\rightarrow{}$ $^2$P$_{1/2}$ transition; the smaller branching ratio can be offset through Purcell enhancement from an optical cavity integrated into an ion trap \cite{Takahashi2020zz}. Using a picosecond pulsed laser at 777\,nm  we generate a second high-intensity pump field at 977\,nm through seeded FWM in photonic crystal fiber (PCF), removing the requirement for a dual pump laser system. Thus, with a single pump laser we drive a larger frequency shift than typically seen in BS-FWM and connect two spectral regions that would require a pump at 3815\,nm to link by TWM\,\cite{PhysRevApplied.7.024021}. Figure\,\ref{fig:1} shows the protocol we employ. PCF\,1 is dispersion-engineered for degenerate FWM with sideband detuning, $\Delta\omega$, equal to the separation between the strontium- and telecommunication-wavelength fields. Either of the FWM sidebands along with light from the initial laser pulse can then be used to drive BS-FWM conversion in PCF\,2. Seeding the first FWM process with a continuous-wave (CW) diode laser at either the signal or idler wavelength limits the spectral width of both sidebands and increases pulse stability\,\cite{Mosley:2011iw}.

\begin{figure}
\centering
{\includegraphics[width=\linewidth]{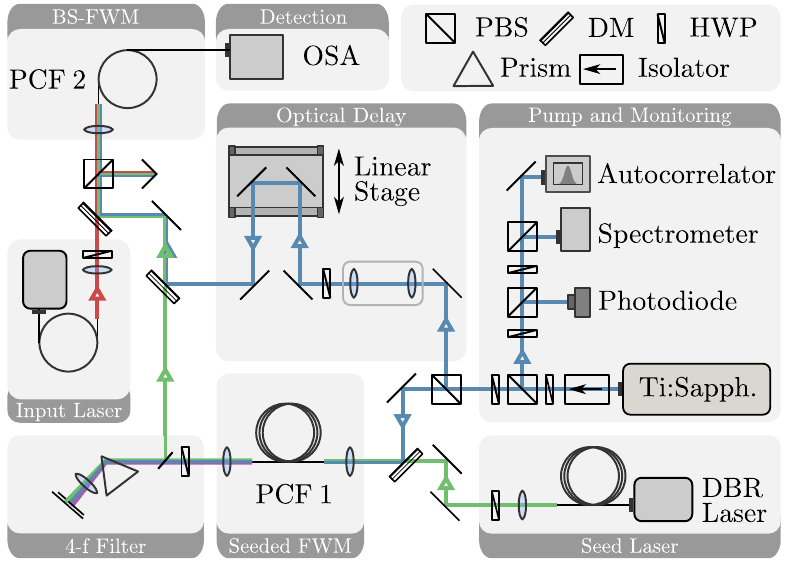}}
\caption{Schematic showing the experimental setup. A full description is given in the text.}
\label{fig:2}
\end{figure}

A schematic of the experimental setup is shown in Fig.\,\ref{fig:2}. The polarisation and intensity of the light from an 80\,MHz repetition rate Ti:Sapphire laser (Spectra-Physics Tsunami) operating at 777-nm wavelength and approximately 12-ps pulse duration were set using half-wave plates (HWPs) and a polarising beamsplitter (PBS). The reflected arm of the PBS was used as a pick-off to monitor the pump pulses with an autocorrelator, spectrometer and Si photodiode. The pump pulses were then split at a second PBS, with the reflected arm sent to a free-space optical delay and the transmitted light ($\sim60\,\%$) directed towards PCF\,1 to pump the seeded FWM. A fiber-coupled 5-MHz-linewidth distributed-Bragg-reflector (DBR) laser with a central wavelength of 977.2\,nm (Thorlabs DBR976P) was used as the CW seed for the FWM in PCF\,1. After the seed-laser polarisation was set to match that of the pump, it was combined with the pump beam on a dichroic mirror and directed towards a 45-cm length of PCF\,1. Both beams were coupled into the fiber using an aspheric lens, which was selected to maximize the coupling efficiency of the pump field. The out-coupling lens which was located after the polarisation-maintaining fiber connected to the diode laser was then selected to maximize the coupling of the seed light in to PCF\,1.

After PCF\,1, the pump, signal, and idler were directed towards a 4-f prism filter to select only the idler field at 977\,nm. The idler was then combined using a dichroic mirror (DM) with the light from the Ti:Sapphire laser which had bypassed PCF\,1 in the optical delay arm, providing 777\,nm pump pulses unaffected by dispersion and nonlinearity in PCF\,1. A Keplerian telescope was available to adjust the beam diameter of the 777-nm pump and a motorized translation stage was used to maximize the temporal overlap between the pulses. The combined BS-FWM pump fields were then mixed at a DM with a source of light at either the Sr$^+$ wavelength or the telecommunication C band and directed towards PCF\,2. An aspheric lens was selected to maximize the coupling efficiency of the 977\,nm pump into PCF\,2 although this was chosen in combination with the collimating lenses for the fiber-coupled C-band laser and  PCF\,1, and the lenses of the Keplerian telescope, to maximize the overall coupling efficiency for all of the fields. The output end of PCF\,2 was inserted in to a bare-fiber adaptor and connected to either an optical spectrum analyser (OSA, Yokogawa AQ6374) or a power meter. Both PCFs used in this experiment were fabricated at the University of Bath by the stack-and-draw technique.

\begin{figure}
\centering
\includegraphics[width=\linewidth]{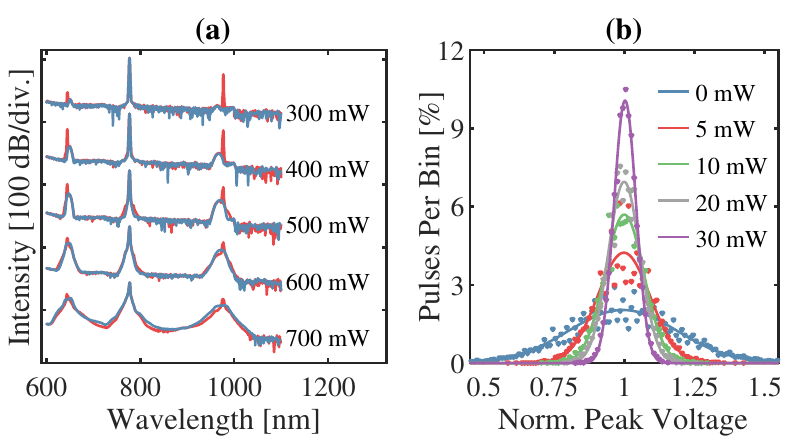}
\caption{(a) Seeded (red) and spontaneous (blue) FWM spectra for increasing pump power. The spectra are offset in increments of 90\,dB for visibility. The spectral resolution is 1\,nm. (b) Distributions of pulse energy in FWM idler pulses for different seed powers. Lines show the least-square Gaussian fits.}
\label{fig:3}
\end{figure}

We first characterized the seeded FWM in PCF\,1. In this fiber, the walk-off length between 12-ps pulses centred at the pump and signal (idler) wavelengths was calculated to be 1.54 (1.37)\,m. The experiment reported here used a 45-cm length of PCF\,1, which we selected for optimum FWM sideband generation; shorter lengths of fiber resulted in less efficient FWM, whereas longer fibers increased nonlinear broadening of the pump and sideband spectra. The spontaneous- and seeded-FWM spectra measured with the OSA are presented in Fig.\,\ref{fig:3}(a). The average pump powers vary between 300 and 700\,mW, while the maximum available seed power is 30\,mW as measured at the output of PCF\,1. Seeding the FWM process serves to significantly increase the sideband spectral amplitudes, but also to reduce the FWHM bandwidths by as much as 77\,\%.  

A pair of Si photodiodes with rise times of 1\,ns were used to analyse the pulse-to-pulse amplitude fluctuations of the seeded FWM. One photodiode measured the intensity of the pulse train emitted by the Ti:Sapphire pump laser, and another was positioned to monitor the idler field transmitted through the 4-f prism filter. The maximum voltage registered by each photodiode for each of $\geq$2000 peaks was recorded using a digital oscilloscope; as the response time of the photodiodes were slow relative to the optical pulse duration, the recorded peak voltages are proportional to the pulse energy.

Figure.\,\ref{fig:3}(b) shows normalized histograms of idler field pulse energy for a fixed pump power of 700\,mW. Without seeding, it is apparent that the spontaneous FWM idler field suffered from significant amplitude noise, with a recorded fractional standard deviation in pulse energy of 20\%. For comparison, the amplitude noise level measured for the pump pulses before the fiber was 1.4\,\%. When the seed power was increased the FWM sidebands exhibited a significantly reduced level of amplitude noise relative to the spontaneous sidebands. The fractional standard deviation in idler pulse energy was always below 10 \% when seeded, decreasing to 4\,\% at the maximum available seed power of 30\,mW. Hence stimulating the FWM process with a seed more than four orders of magnitude lower in peak power than the pump results in a remarkably stable, high intensity idler suitable to be used as a pump for BS-FWM. We expect that, with modest additional seed power, this amplitude noise could be reduced even further. For BS-FWM conversion, we operated the seeded FWM at a pump power of 650\,mW and a seed power of 30\,mW. In this configuration the idler had an average power of 45\,mW after the 4-f filter and a standard deviation in pulse energy of 3.5\,\%; the FWHM spectral bandwidth was $<$2\,nm.

\begin{figure}[ht]
\centering
{\includegraphics[width=\linewidth]{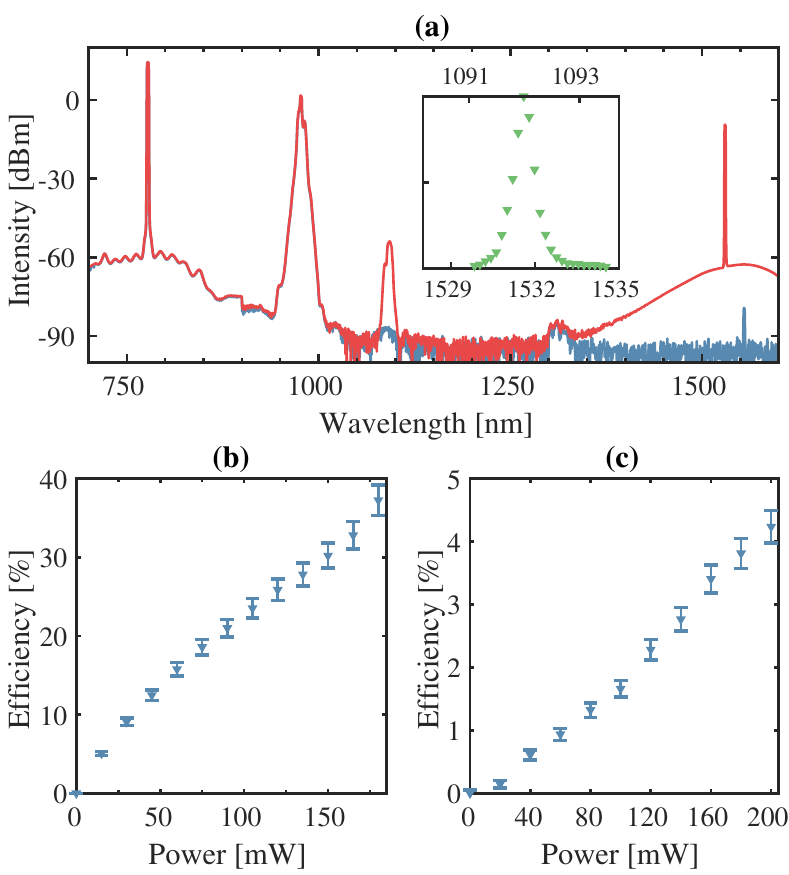}}
\caption{(a) BS-FWM spectra from a 1.2-m length of PCF\,2 with pump powers of 125\,mW and 12.5\,mW at 777\,nm and 977\,nm, respectively. Blue: C-band source blocked. Red: C-band source unblocked, showing conversion to 1092\,nm. The spectral resolution is 2\,nm. Inset: The input-wavelength-dependant conversion efficiency of BS-FWM in PCF\,2. The wavelength of the input and converted fields are shown on the lower and upper x axes, respectively. (b) The 777-nm-pump-power-dependant conversion efficiency of BS-FWM in a 1.2-m length of PCF\,2. The average power of the 977-nm pump was 18\,mW. (c) The 777-nm-pump-power-dependant conversion efficiency of BS-FWM down-conversion in a 1.2-m length of PCF\,2. The average power of the 977-nm pump was 17\,mW.}
\label{fig:4}
\end{figure}

We first interrogated the performance of BS-FWM in PCF\,2 through the down-conversion of coherent CW light from a fiber-coupled wavelength-tuneable sub-100-kHz-linewidth laser in the telecommunication C band (ID Photonics CoBrite DX1). Typical output spectra at the output of PCF\,2 are shown in Fig.\,\ref{fig:4}(a).  Without an input at the source telecoms wavelength, we see the two pump wavelengths at 777\,nm and 977\,nm along with a number of parasitic nonlinear processes, including a small signal at 1316\,nm arising from non-phase-matched degenerate FWM pumped at 977\,nm and seeded at 777\,nm as well as another at 1554\,nm, the second harmonic of the 777\,nm pump. With an input field at 1530\,nm, we see a clear frequency-converted peak at the target wavelength of 1092\,nm. Note that the fraction of the CW input field that overlaps temporally with the pump pulses was approximately $10^{-3}$ reducing the displayed spectral intensity of converted to input light by 30\,dB. 

We collected spectra on the OSA and integrated power at the Sr$^+$ and telecommunication wavelengths to estimate the BS-FWM conversion efficiency. 
We calculate the internal up-conversion efficiency normalized to mean photon number, $\eta_{\mathrm{up}}$, according to 
\begin{equation}
	\eta_{\mathrm{up}} = \frac{(P_{\mathrm{\,Sr}}-N_{\mathrm{\,Sr}})\,\omega_{\mathrm{T}}}{P_{\mathrm{\,T\,}}D\,\omega_{\mathrm{Sr}}},	
\end{equation}
where $P_{\mathrm{\,T}}$ is the integrated power at the input telecommunication wavelength, and the values of $P_{\mathrm{\,Sr}}$ and $N_{\mathrm{\,Sr}}$ are the integrated powers at the Sr$^+$ wavelength with and without the telecommunication input field blocked. The duty cycle, $D$, is determined by the repetition rate, $R_{\mathrm{P}}$, and pulse duration, $\uptau _{\mathrm{p}}$, of the pumps fields; $D = \uptau _{\mathrm{p}}R_{\mathrm{P}}$.

We characterized the bandwidth of BS-FWM phase matching in a 1.2-m length of PCF 2 by tuning the telecommunication wavelength input field and recording the resultant conversion efficiency for fixed pump powers of 14\,mW at 977\,nm and 30\,mW at 777\,nm.  The results are inset within Fig.\,\ref{fig:4}(a), showing a phase-matching peak which is centred at 1531.6\,nm and has a FWHM bandwidth of 0.9\,nm. The maximum up-conversion efficiency we observed was 37$\pm$2\%, which was achieved using average pump powers of 18\,mW and 180\,mW at 977\,nm and 777\,nm, respectively. As is apparent from Fig.\,\ref{fig:4}(b), the efficiency was limited by the power of the pumps; a longer length of fiber would not increase the efficiency due to the walk-off length for the pump pulses. It is likely that, with an improved selection of lenses and with careful alignment, the amount of  available pump power coupled into PCF\,2 could be increased; in the conversion reported here, only $\sim$40\% of the light produced by the seeded FWM in PCF\,1 was successfully coupled into PCF\,2.

To characterize frequency down-conversion from 1092\,nm to the C-band, we replaced the telecom-wavelength laser with a home-built CW external-cavity diode laser (ECDL) based on a gallium arsenide gain chip from Eagleyard Photonics. 
The BS-FWM down-conversion efficiency was calculated as
\begin{equation}
	\eta _{\mathrm{down}} = \frac{(P_{\mathrm{\,T}}-N_{\mathrm{\,T}})\,\omega_{\mathrm{Sr}}}{P_{\mathrm{\,Sr\,}}D\,\omega_{\mathrm{T}}}.	
\end{equation}
The 777-nm-pump-power-dependant conversion efficiency is shown in Fig.\,\ref{fig:4}(c), from which we observe a maximum conversion efficiency of 4.2$\pm$0.3\% at pump powers of 17\,mW and 200\,mW at 977\,nm and 777\,nm, respectively. Similar to the power-dependant up-conversion shown, no roll off in the efficiency at higher power occurs, indicating the availability of more pump power would increase the performance of the conversion. {Although the bandwidth of the ECDL was narrower than our OSA resolution (0.05\,nm), we believe the possible disparity in frequency-mode structure of the source fields, imbalance in pump powers, and dispersion of nonlinearity contribute to the asymmetry between up- and down-conversion.} Monitoring the count rate in the C-band channel with an InGaAs single-photon counter suggested that noise as a result of the two pump fields is sufficiently low for operation in a quantum network \cite{Wright:2020th}.

Within an ion-trap quantum network there are temporal, spectral bandwidth, and polarisation-stability requirements which need to be met by any frequency converter\,\cite{Bock:2018jw,Walker:2018jh}.
Our scheme can be readily adapted with commercially-available lasers that achieve similar peak powers in the nanosecond regime but match the duration and bandwidth of trapped-ion emission, maintaining conversion efficiency. Alternatively, when shifting from the picosecond to nanosecond regime, longer lengths of fiber could be used as the walk-off lengths for the pulses are three orders of magnitude larger.

In summary, we have presented a resource-efficient BS-FWM conversion scheme which achieves frequency conversion with a single pulsed pump laser by sequential use of  FWM. We have demonstrated bi-directional conversion of coherent light between a Sr$^+$ emission wavelength at 1092\,nm and the telecommunication C band, using a pump laser with picosecond pulses with up- and down-conversion efficiencies of 37$\pm$2\% and 4.2$\pm$0.3\%, respectively. This conversion scheme may be applied to ion-trap quantum networks by using nanosecond duration pump pulses. 

Data underlying the results presented are available at \cite{bata546}. 

\medskip
\noindent\textbf{Funding.} This work was funded by the UK EPSRC Quantum Technology Hub NQIT (EP/M013243/1) and the Quantum Computing and Simulation Hub (EP/T001062/1).
\medskip

\noindent\textbf{Disclosures.} The authors declare no conflicts of interest.

\bibliography{references}

\end{document}